\documentclass{PoS}

\title{Non-custodial warped extra dimensions at the LHC}

\ShortTitle{Non-custodial warped extra dimensions at the LHC}

\author{Barry M. Dillon\\
        University of Sussex\\
        E-mail: \email{bd91@sussex.ac.uk}}

\author{Stephan J. Huber\\
        Univeristy of Sussex\\
        E-mail: \email{s.huber@sussex.ac.uk}}

\abstract{With the prospect of improved Higgs measurements at the LHC and at proposed future colliders such as ILC, CLIC and TLEP we study the non-custodial Randall-Sundrum model with bulk SM fields and compare brane and bulk Higgs scenarios.  We compute the electroweak precision observables and argue that incalculable contributions to these, in the form of higher dimensional operators, could have a non-negligable impact.  This could potentially reduce the bound on the lowest Kaluza-Klein gauge boson masses to the $5$ TeV range, making these states detectable at the LHC.  In a second part, we compute the misalignment between fermion masses and Yukawa couplings caused by vector-like Kaluza-Klein fermions.  The deviation of the top Yukawa can easily reach $10\%$, making it observable at the high-luminosity LHC. Corrections to the bottom and tau Yukawa couplings can be at the percent level and detectable at ILC, CLIC or TLEP.}

\FullConference{18th International Conference From the Planck Scale to the Electroweak Scale \\
		25-29 May 2015\\
		Ioannina, Greece }

\begin{document}

\section{Introduction}
Due to their attractive model building features and rich phenomenology, warped extra dimensional models have been studied extensively for fifteen years.  The first proposal of such a model was by Randall and Sundrum (RS) in 1999 \cite{Randall:1999ee}, and consisted of an AdS space mapped onto an $S_1/Z_2$ orbifold bounded by two 3-branes.  The AdS geometry imposes an exponential hierarchy in energy scales between the two branes, thus, with all standard model (SM) fields residing on the low energy (IR) brane and with a suitable choice of parameters, this model offers a simple and natural solution to the hierarchy problem.

Putting the standard model (SM) fields in the bulk introduces towers of new states, known as Kaluza-Klein (KK) modes.  The zero modes of these are associated with the SM states.  The mixing between the KK modes and the zero modes modifies the couplings and masses in the low energy effective theory, and thus provides constraints on the 5D model.  Currently, some of the largest constraints come from the electroweak precision observables (EWPOs).  In models without a custodial symmetry, the EWPOs put a lower bound on the lightest spin-1 resonances of $\sim 8$ TeV.  In this paper we study incalculable effects accounted for by higher dimensional operators in the 5D bulk and find that they are, in some cases, non-negligable.  We find that, without fine-tuning, these effects could lower the bound on the lightest spin-1 resonance to $\sim 5$ TeV.

The large coupling between the top and the Higgs leads us to expect large deviations to the standard model top Yukawa coupling.  This deviation arises from mixing with vector-like KK top modes.  Studying the effects of the lowest laying vector-like modes we find that corrections to this coupling are of the order $\sim 10\%$, which could be interesting for LHC searches.  Corrections to other couplings are generally smaller, but are still percent level for the bottom quark and tau lepton.  These percent level corrections are out reach of current colliders, but could be searched for at ILC, CLIC or TLEP.    

\section{Bulk fields in Randall-Sundrum}
The Randall-Sundrum background is defined by the non-factorizable metric \cite{Randall:1999ee}:
\begin{equation}
ds^2=e^{-2k|y|}\eta_{\mu\nu}dx^{\mu}dx^{\nu}-dy^2 .
\end{equation}
The 4D metric is $\eta_{\mu \nu}=$diag$(1,-1,-1,-1)$, $k$ is the AdS curvature, and $y$ defines the position along the extra dimension.  The extra dimension is bounded by two 3-branes in the UV ($y=0$) and in the IR ($y=L$).  The length of the extra dimension, $L$, is assumed to be O$(11 \hspace{1mm} \pi k^{-1})$ and is the free parameter which determines the new physics scale.  The AdS curvature can be expressed in terms of the fundamental 5D mass scale $M_5$ by
\begin{equation}
\kappa = \frac{k}{M_5}.
\end{equation}
This is discussed in more detail in \cite{Davoudiasl:1999jd} where the authors use $0.01\leq\kappa\leq1$ for their phenomenological analysis.  However, at the larger end of this range higher derivative corrections to the gravitational action will become important, rendering the derivation of the metric (1) unreliable.

To treat the Higgs in the bulk we need to look at scalar fields in 5D.  After expanding in Kaluza-Klein modes,
\begin{equation}\label{KKscalar}
\Phi(x,y)=\frac{1}{\sqrt{L}}\sum_n \Phi_{n}(x)f_n(y)
\end{equation}
we find that, assuming we have the correct boundary conditions, a massless zero mode exists with the 5D profile,
\begin{equation} \label{scalarzero}
f_0(y)=\sqrt{\frac{2(1+a^2)kL}{1-e^{-2(1+a^2)kL}}}e^{-a^2ky}.
\end{equation}
The  parameter $a^2$ defines the localisation of the field in 5D and $a^2<0$ implies IR localisation.  Along with this zero mode one obtains a tower of IR localised KK scalar fields with masses O(TeV).  In order to solve the hierarchy problem, i.e. for O($M_{Pl}$) 5D mass terms to get suppressed to O(TeV), we require $a^2\leq -2$.

For gauge fields the treatment is similar, except that no bulk or brane potential terms are allowed.  This means that, in the absence of EWSB, there is a flat massless mode in the spectrum.  Here also there are a tower of IR localised KK states with masses O(TeV).

The treatment of fermions is complicated slightly by the fact that 5D Dirac fermions are not chiral.  For a detailed discussion of fermions in 5D see \cite{Csaki:2005vy,Gherghetta:2010cj}.  Lorentz invariance implies that we must work with full Dirac spinors in 5D.  We can project out one of the zero mode Weyl spinors, leaving us with chiral zero modes and towers of vector-like KK modes.  A remarkable property of fermions in 5D is that we can add a mass to the action while keeping a massless chiral zero mode in the spectrum.  The effect of this mass term is that it allows us to exponentially localise the zero mode anywhere in the bulk, while the KK states remain IR localised.  The zero mode profile can be written as,
\begin{equation} \label{fermionzero}
f_{\pm}^{(0)}(y)=\frac{1}{N^{(0)}_{\pm}}e^{(2\mp c)ky}
\end{equation}
where $c$ comes from the 5D mass $m_{\Psi}=ck$ and $N^{(0)}_{\pm}$ is a normalisation constant.  The plus and minus refer to left and right handed chiralities, respectively.  This exponential localisation allows us to generate a hierarchy in fermion Yukawa couplings with order one mass terms in 5D.  For a more detailed discussion of these bulk fields, see \cite{Dillon:2014zea,Ahmed:2015ona}.

\section{Electroweak precision observables}
\subsection{Calculable Corrections}
We consider a non-custodial $SU(2)_L\times U(1)_Y$ bulk gauge sector with bulk fermions and a bulk Higgs.  Calculating the Peskin-Takeuchi parameters \cite{Peskin:1991sw} is straightforward, and assuming universal UV fermion localisations for the light fermions, we can account for their effects also.  Corrections to the SM can arise from the zero mode fields mixing with KK modes, and from the exchange of KK particles in a physical process. For our purposes the latter is only a small effect and will be ignored.  For a detailed analysis of the case of a brane Higgs see e.g.~\cite{Delgado:2007ne}.  

In general, the $T$-parameter measures corrections to the standard model ratio $m_Z/m_W$.  In our model, tree-level corrections arise via mixings between gauge KK modes induced by the Higgs vev.  We parameterise this mixing with $R_n={M_{0n}^2}/{v_0^2}$ where,
\begin{equation}
M_{mn}^2=\frac{v_0^2}{L}\int_0^Ldy\hspace{1mm}e^{-2ky}w_mw_nf_0^2
\end{equation}
where $w_m$ is the profile of the $m^{th}$ gauge mode and $f_0$ is the vev profile.  This mixing induces shifts in the gauge couplings to fermions in the effective action.  The effect of this shift is encoded in the $S$-parameter and will thus depend on the coupling between the zero mode fermions and the gauge KK modes,
\begin{equation}
g_{0n}=\frac{g_5}{L^{\frac{3}{2}}}\int_0^Ldy\hspace{1mm}e^{-3ky}(f_{+}^{(0)})^2w_n.
\end{equation}
The $U$-parameter in this model doesn't receive corrections at tree-level, so we will not discuss it any further.  With the above expressions we can now write down the tree-level expressions for the electroweak parameters,
\begin{equation} \label{S}
S\simeq\left(\frac{-9\pi}{2}\sum_n\frac{R_n}{(n-\frac{1}{4})^2}\frac{g_{0n}}{g_{00}}\right)\frac{v_0^2}{M_{KK}^2}
\end{equation}
\begin{equation}\label{T}
T\simeq\left(\frac{9\pi}{16c_W^2}\sum_n\frac{R_n}{(n-\frac{1}{4})^2}\left(R_n+2\frac{g_{0n}}{g_{00}}\right)\right)\frac{v_0^2}{M_{KK}^2} \\
\end{equation}
where $M_{KK}$ the mass of the first gauge boson excitation and $v_0$ is the standard model vev.  Neglecting contributions from higher KK modes, we find a correlation between the $S$ and $T$ parameters which can be expressed as
\begin{equation}\label{STcorr}
T\simeq\frac{1}{8c_W^2}\left(2-\frac{g_{00}}{g_{01}}R_1\right)S.
\end{equation}
Depending on $T/S$, the model can live in more or less experimentally favoured regions of the parameter space, possibly resulting in reductions to the $M_{KK}$ constraint.   

Table \ref{higgstable} shows that the bulk Higgs couples less to gauge KK modes than the brane Higgs.  As a result, not only will the $T$ parameter be smaller for a bulk Higgs, but we find that a two mode approximation is sufficient for a bulk Higgs, but not for a brane Higgs.
\begin{table}[t]
\centering
\begin{tabular}{lllll}
 & $R_1$ & $R_2$ & $R_3$ & $R_4$ \\
Brane Higgs & 8.4 & -8.3 & 8.1 & -8.2 \\
Bulk Higgs ($a^2=-2$) & 5.6 & -0.9 & 0.5 & -0.3 
\end{tabular}
\caption{Here we show how the couplings between the zero mode Higgs and the gauge KK tower differ for a brane and bulk Higgs.}
\label{higgstable}
\end{table}
Light fermions must be localised in the UV so that their overlap with the Higgs is small, this corresponds to $c_L>0.5$ \cite{Grossman:1999ra,Gherghetta:2000kr,Huber:2003tu} where ~$g_{0n}/g_{00}\lesssim0.2$.  We can see that this implies a small coupling with the KK gauge modes and therefore small vertex contributions in the electroweak parameters.
\newline
\newline
Current bounds on $S$ and $T$ with $U=0$ are given in \cite{Baak:2012kk} (see figure \ref{figurest}). Taking the 95\% CL bound, we find the following bounds for a brane and bulk Higgs:
\begin{itemize}
\item \underline{Brane Higgs}:  Due to the large values of $R_n$ the KK gauge modes have large contributions to the $T$ parameter.  If we approximate $R_n^2\simeq8.4^2$ for all $n$, we can sum the full tower contributions by taking the sum $\sum_{n=1}^{\infty}(n-1/4)^{-2}\simeq2.54$.  We then find that the electroweak constraints require $M_{KK}\gtrsim15$ TeV.
\item \underline{Bulk Higgs} ($a^2=-2$):  Since the $R_n$ values are small for $n>1$ we find that the first mode makes the only sizeable contribution to the electroweak parameters.  With just the first excited mode we find the bounds to be $M_{KK}\gtrsim8$ TeV.  Including the first 10 modes only corrects the $8$ TeV bound by $0.3\%$, and the second excited mode contributes $0.26\%$ of this correction.  We find similar effects for the $S$ parameter.
\end{itemize}
These results are in agreement with the bounds found elsewhere in the literature \cite{Cabrer:2011fb,Archer:2012qa,Fichet:2013ola,Archer:2014jca,Ahmed:2015}.

Another thing one should consider is the deviation in the standard model gauge boson masses and their coupling to the Higgs zero mode.  We find that the HHZ and the HHZZ interactions receive identical corrections $\sim R_1^2\hspace{1mm}m_Z^2/M_{KK}^2$,  and similarly for the W boson.  With the lightest gauge boson mass at $8$ TeV we find a $0.4\%$ misalignment for the Z boson and a $0.3\%$ misalignment for the W boson.  This would be visible at the ILC \cite{Peskin:2012we,Asner:2013psa} or TLEP \cite{Gomez-Ceballos:2013zzn}.  The only way to reduce this misalignment is to either increase $M_{KK}$ or to reduce the coupling of the Higgs zero mode to the gauge KK modes, which can be achieved by modifying the background geometry in the bulk \cite{Falkowski:2008fz,Cabrer:2011fb,Carmona:2011ib}.  

\subsection{Higher dimensional operator contributions to $S$, $T$ and $U$}
In the previous section we demonstrated how to estimate the size of the calculable contributions to the electroweak parameters in the 4D effective theory.  There will also be incalculable contributions from the UV theory which we will parameterise using higher dimensional operators in the 5D theory.  The three leading operators contributing to the oblique parameters are
\begin{equation}\label{STU}
S: \hspace{3mm} \frac{\rho}{M_5^3}(\Phi^{\dagger}T^a\Phi)W^a_{MN}B_{MN}, \hspace{5mm} T:  \hspace{3mm} \frac{\lambda}{M_5^3}|\Phi^{\dagger}D^M\Phi|^2, \hspace{5mm} U: \hspace{3mm} \frac{\theta}{M_5^6}|\Phi^{\dagger}W^{MN}\Phi|^2,
\end{equation}
where $\rho$, $\lambda$ and $\theta$ are unknown parameters. These operators could be present both on the branes or in the bulk, i.e.~$\rho=\rho_B+\rho_{IR}M_5^{-1}\delta(y-\pi R)$.  In the brane case there is an extra mass scale suppression. There is also a possible contribution from the UV brane, which is negligible for an IR localised Higgs.

\begin{figure}[t]
\centerline{\includegraphics[scale=0.25]{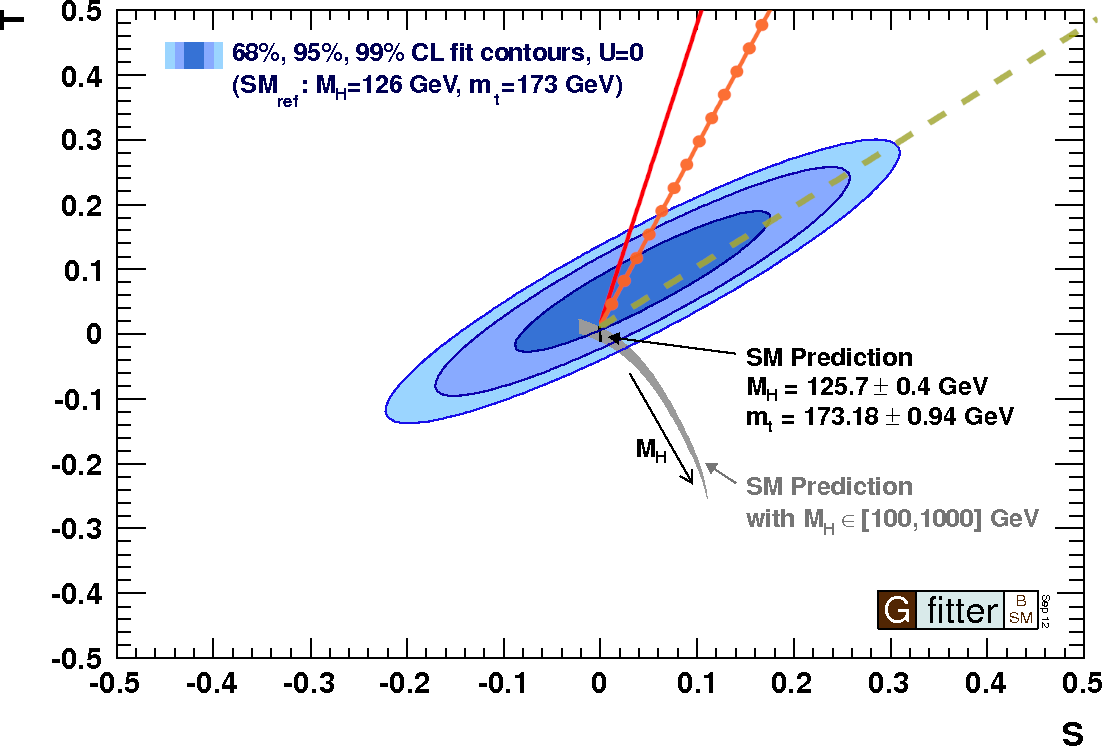}}
\caption{Here we have overlaid the bounds from \cite{Baak:2012kk} with the $S$ and $T$ correlations for $\delta_6=0$ (solid), $\delta_6=-0.4$ (dots) and $\delta_6=-0.8$ (dashed).}
\label{figurest}
\end{figure}

The $S$ and $T$ operators both have effective coefficients $\sim v_0^2/M_{KK}^2$, but due to the higher dimension of the $U$ operator it is of the order $\sim \left(v_0^2/M_{KK}^2\right)^2$.  This behaviour in the $U$ parameter has been noted in \cite{Barbieri:2004qk} also.  Thus only $S$ and $T$ will receive sizeable corrections from these operators, while $S$ also has an additional suppression $\sim \frac{1}{kL}$ with the respect to the $T$ coefficient.  All three operators show similar dependence on the Higgs localisation. The effective coefficients grow exponentially as $a^2$ decreases until, at $a^2=-1$, the exponential growth stops, which is due to the normalisation of the Higgs field.  At $a^2<-1$, operators on the IR brane increase linearly with $a^2$ while operators in the bulk remain mostly constant.  However, to ensure we have the correct suppression of UV mass perturbations, we keep $a^2=-2$.  Assuming O($1$) values for the coefficients of the higher dimensional operators in eq.~\ref{STU}, we would only expect a sizeable contribution from the operator contributing to the $T$-parameter.  If we ignore the vertex corrections, and include the effects of these operators, we find the total $T$ parameter
\begin{equation}
T_6\simeq\left(\frac{3\pi}{4}\right)^2\left(\frac{1}{\pi c_W^2}\sum_n\frac{R_n^2}{(n-0.25)^2}+\frac{2}{3}\frac{\kappa^3}{\alpha}\lambda_B+4\frac{\kappa^4}{\alpha}\lambda_{IR}\right)\frac{v_0^2}{M_{KK}^2}=T\left(1+\delta_6\right),
\end{equation}
where we again take $M_{KK}=m_1\simeq(3\pi/4)ke^{-kL}$ and the $B$ and $IR$ subscripts refer to bulk and brane contributions to the higher dimensional operators.
Here $\delta_6$ parameterises the contribution from higher dimensional operators,
\begin{equation}
\delta_6=\left(\frac{1}{\pi c_W^2}\sum_n\frac{R_n^2}{(n-0.25)^2}\right)^{-1}\left(\frac{2}{3}\frac{\kappa^3}{\alpha}\lambda_B+4\frac{\kappa^4}{\alpha}\lambda_{IR}\right).
\end{equation}
For $\lambda_{IR}=\lambda_{UV}=1$ the contributions from $\delta_6$ are small, of order $\sim 0.1$ for $\kappa\sim0.2$.  Whereas for $\lambda_{IR}=\lambda_{UV}=10$ the contributions are of order $\sim 0.7$ for $\kappa\sim0.2$.  Hence large corrections are possible.  These corrections also modify eq.~(\ref{STcorr}) such that the correlation is expressed as
\begin{equation}
T_6\simeq\frac{1}{8c_W^2}\left(2-\frac{g_{00}}{g_{01}}R_1\right)\left(1+\delta_6\right)S.
\end{equation}
From figure \ref{figurest} we see that as well as directly reducing the $T$ parameter, $\delta_6\neq0$ can take us to a more favourable region of the S-T plane, depending on the relative sign.  This would allow for a further reduction on the $M_{KK}$ bound.  If we take the 95\% CL bound from figure 4, we find that the lower bound on $M_{KK}$ is approximately $6$ TeV and $2.7$ TeV for $\delta_6=-0.4$ and $-0.8$, respectively. So it is plausible to assume that incalculable contributions to the $T$ parameter lead to a partial cancellation and so relax the bound on the KK scale. It therefore seems premature to exclude discovery of such a scenario at the forthcoming LHC run.

\section{Yukawa coupling corrections}

In the same way the Higgs vev induced mixing among gauge KK modes, it also induces mixing among the fermion KK modes.  However the fermion case is complicated slightly due to the vector-like structure of the KK states.  The mass matrix for the top quark, after EWSB, is of the form,
\begin{equation} \label{fermionmassm}
\left(\begin{array}{cccc}\bar{T}_L^0 & \bar{T}_L^1 &\bar{t}_L^1 &\ldots \end{array}\right)
\left(\begin{array}{cccc}         m^{T,0}_{t,0} & 0 & m^{T,0}_{t,1} & \ldots \\ m^{T,1}_{t,0} & M_{T,1} & m^{T,1}_{t,1}  & \ldots \\ 0 & m^{t,1}_{T,1} &M_{t,1}  &\ldots \\  \vdots & \vdots&\vdots  & \ddots          \end{array}\right)
\left(\begin{array}{c} t_R^0 \\ T_R^1 \\ t_R^1\\  \vdots \end{array}\right),
\end{equation}
where $t$ is an $SU(2)$ singlet and $T$ is one component of the $SU(2)$ doublet $Q=(T,B)$.  For $t$ we keep only the right handed zero mode and for $T$ we keep only the left handed zero mode, while both fields have vector-like KK towers where the super-scripts denote the state.  The diagonal elements of the mass matrix are simply the KK masses gained from the 5D theory, while the off-diagonal elements are the mixings induced by the Higgs.  Note that these do not overlap, like in the gauge case.  The mixings are given by
\begin{equation}
m^{\psi ,m}_{\phi ,n}=\frac{1}{\sqrt{2}}\lambda^{\psi ,m}_{\phi ,n}v_0=\frac{\lambda_t^{(5)} v_0}{\sqrt{2}L}\int_0^L dy \hspace{1mm} \sqrt{|g|} f^m_{\psi L} f^n_{\phi R} f_0,
\end{equation}
where $f^m_{\psi L}$ and $f^n_{\phi R}$ are the 5D fermion profiles, $f_0$ is the Higgs profile and $\lambda_t^{(5)}$ is the 5D Yukawa coupling.  Neglecting CP violation, the mass matrix (\ref{fermionmassm}) can be partially diagonalized using orthogonal transformations of the left and right handed KK modes.  This transformation will isolate the ``zero mode'' from the KK excitations.  Since the mixings are $\sim v_0^2/M_{KK}^2$ we can assume a small angle approximation.  For the mass of the lowest lying mode (``zero mode'') we then find 
\begin{equation}\label{mT}
m^{(4)}_t=m^{T,0}_{t,0}\left(1-\frac{(m^{T,0}_{t,1})^2}{2M_{t,1}^2}-\frac{(m^{T,1}_{t,0})^2}{2M_{T,1}^2}+\left(\frac{m^{t,1}_{T,1}}{m^{T,0}_{t,0}}\right)\frac{m^{T,0}_{t,1}m^{T,1}_{t,0}}{M_{T,1}M_{t,1}}+O\left(\frac{m^3}{M_{KK}^3}\right)\right).
\end{equation} 
A matrix analogous to eq.~(\ref{fermionmassm}) encodes the Yukawa interactions of the fermion KK modes with the Higgs. In this matrix, diagonal terms corresponding to $M_{T,n}$ and $M_{t,n}$ are missing. This  results in a relative shift between the Yukawa coupling and mass of the ``zero mode'' compared to the standard model. We find that the ``zero mode'' Yukawa coupling can be written as
\begin{equation}
\lambda^{(4)}_t=\lambda^{T,0}_{t,0}\left(1-\frac{3}{2}\frac{(\lambda^{T,1}_{t,0}v_0)^2}{M_{T,1}^2}-\frac{3}{2}\frac{(\lambda^{T,0}_{t,1}v_0)^2}{M_{t,1}^2}+3\left(\frac{\lambda^{t,1}_{T,1}}{\lambda^{T,0}_{t,0}}\right)\frac{\lambda^{T,0}_{t,1}\lambda^{T,1}_{t,0}v_0^2}{M_{T,1}M_{t,1}}+O\left(\frac{\lambda^3v_0^3}{M^3_{KK}}\right)\right).
\end{equation}
We can now quantify the misalignment in the fermion ``zero mode'' mass and Yukawa coupling as
\begin{equation} \label{rt}
r_t^{(4)}=\frac{\sqrt{2}\hspace{0.5mm}{m}^{(4)}_{t}}{{\lambda}^{(4)}_{t}v}-1=\frac{(\lambda^{T,1}_{t,0}v_0)^2}{M_{T,1}^2}+\frac{(\lambda^{T,0}_{t,1}v_0)^2}{M_{t,1}^2}-2\left(\frac{\lambda^{t,1}_{T,1}}{\lambda^{T,0}_{t,0}}\right)\frac{\lambda^{T,0}_{t,1}\lambda^{T,1}_{t,0}v_0^2}{M_{T,1}M_{t,1}}+ \frac{\delta w}{2} +O\left(\frac{\lambda^3v_0^3}{M^3 _{KK}}\right).
\end{equation}
Note that because $\lambda^{T,0}_{t,1}$ is negative there is no cancellation in the contributions to $r_t^{(4)}$.
The $\delta w$ term is related to the gauge boson mass correction from KK mixing. We use it here because the measured vacuum expectation value., $v$, includes the mass correction to the $W$ boson.  We only include this factor for completeness since from the electroweak precision tests we know that it does only result in a negative per-mille  correction. 

Our numerical evaluations of   $r_t^{(4)}$ are summarised in table \ref{results}. For the case of a bulk Higgs we use a KK scale of $M_{KK}=5.9$ TeV. 
As discussed in the previous section, a small contribution from higher dimensional operators is required in this case to reduce electroweak constraints to meet experimental bounds. For a KK scale of 8 TeV, the Yukawa deviations from the table will be reduced by a factor of $(5.9/8)^2=0.54$, while for a KK scale of 5 TeV they will increase by a factor of 1.4.
We give separate results for the three individual contributions and the total result from eq.~(\ref{rt}), $r_t^{(4)}$, denoted by (a), (b), (c) and Total, respectively. As anticipated, the third term (c) is always very important, and completely dominates for smaller fermion masses. Note that the scaling in 5D Yukawa couplings is somewhat distorted by changes in the fermion locations needed to keep the fermion mass constant. For the top quark these modifications can easily be larger than the anticipated 4\% accuracy from HL-LHC \cite{CMS:2013xfa,ATLAS:2013hta}. Also for the bottom quark the correction in the Yukawa coupling could be larger than the 2.4\% or 0.4\% accuracies aimed for at ILC and TLEP, respectively \cite{Gomez-Ceballos:2013zzn}. For the tau Yukawa coupling it seems questionable whether a deviation could be seen at ILC (predicted accuracy 2.9\%), while a detection at TLEP (predicted accuracy 0.5\%) seems promising \cite{Gomez-Ceballos:2013zzn}. 

\begin{table}[ht]
\centering
\begin{tabular}{cccccccccc}
\hline
$m_t^{(4)}$ & $\lambda_t^{(5)}$ & $c_L$ & $c_R$ & $M_{T1}$ & $M_{t1}$ & (a) & (b) & (c) & Total 
\\ 
$[\rm GeV]$ & & & & [TeV] &[TeV] & [\%] & [\%] & [\%] & [\%] \\
\hline
&&&&&&&&& \\
173.48 & 4 & 0.550 & -0.26 & 6.52 & 7.12 & 12.97 & 0.05 & 19.72 & 32.7 \\
173.73 & 2 & 0.530 & -0.07 & 6.05 & 7.64 & 5.93 & 0.01 & 3.35 & 9.29 \\
173.07 & 1 & 0.488 & -0.20 & 5.98 & 7.12 & 1.29 & 0.03 & 1.31 & 2.62 \\ 
&&&&&&&&&\\ 
4.17 & 4 & 0.526 & -0.6320 & 6.04 & 6.46 & $\sim10^{-3}$ & 0.02 & 6.76 & 6.78 \\
4.17 & 2 & 0.510 & -0.6190 & 5.97 & 6.41 & $\sim10^{-3}$ & 0.02 & 2.48 & 2.50 \\
4.17 & 1 & 0.500 & -0.6004 & 5.93 & 6.33 & $\sim10^{-3}$ & 0.02 & 0.98 & 1.00 \\
&&&&&&&&&\\
1.79 & 4 & 0.542 & -0.650 & 6.10 & 6.53 & $\sim10^{-3}$ & $\sim10^{-3}$ & 3.86 & 3.87 \\
1.79 & 2 & 0.508 & 0.650 & 5.97 & 6.53 & $\sim10^{-4}$ & $\sim10^{-3}$ & 1.07 & 1.08 \\
1.79 & 1 & 0.516 & -0.621 & 6.00 & 6.41 & $\sim10^{-4}$ & $\sim10^{-3}$ & 0.58 & 0.58 \\
\end{tabular}
\caption{Relative shifts in the 4D Yukawa coupling, $r_t^{(4)}$, from eq.~\protect\ref{rt}. The columns denoted by (a), (b), (c) and Total give the first, second, third contribution and the total result in percent.
$M_{KK}$ is taken to be 5.9 TeV.}
\label{results}
\end{table}
We have numerically verified that the expressions (\ref{mT}) to (\ref{rt}), which are derived from considering a single KK level, receive only small corrections of $\lesssim 10\%$ when we include more fermion KK modes.  Yukawa coupling misalignment also has impact on flavour violation mediated by Higgs exchange, as e.g.~discussed in \cite{Casagrande:2008hr,Buras:2009ka,Azatov:2009na}. Also Higgs corrections to the muon anomalous magnetic moment were found to depend on the Higgs localisation \cite{Beneke:2014sta}.
Analysing the resulting constraints for the scenario investigated here, however, is beyond the scope of the present work.

\section{Conclusions}
We have revisited the scenario of a warped extra dimension with a bulk Higgs in the absence of a custodial symmetry or modifications to AdS.  Studying the electroweak precision constraints we confirm the result that a bulk Higgs, rather than a brane Higgs, reduces the lower bound on spin-1 states from $\sim 15$ TeV to $\sim 8$ TeV.  We then consider the impact that bulk higher dimensional operators in the 5D theory has on these electroweak precision observables (EWPOs) and find that the only sizeable contribution comes from the dimension 6 operator contributing to the $T$-parameter.  This contribution is of the same order as the tree-level contribution from the KK mixing.  Therefore we argue that the 8 TeV bound is not robust, and show that with some cancellations the bound could easily be as low as $\sim 5$ TeV.  With this bound, it may be possible that this particular model is testable at the LHC.  KK mixing induced by the Higgs vev also occurs in the fermion sector, and causes shifts in the Yukawa couplings that we would expect from the SM.  We calculate this shift and find that for the top quark the correction is of the order $\sim 10\%$, whereas for the tau lepton and bottom quark the correction is percent level.  With a top Yukawa shift of this size, the LHC may be able to test this model and put bounds on the KK mixing.  For the lighter quarks the ILC or TLEP may have the sensitivity to probe these shifts, however they will be out of reach of the LHC.

\bibliography{BibTeXReferenceList2}{}

\providecommand{\href}[2]{#2}\begingroup\raggedright\begin{thebibliography}{10}

\bibitem{Randall:1999ee}
L.~Randall and R.~Sundrum, {\it A large mass hierarchy from a small extra
  dimension},  {\em Phys.Rev.Lett.} {\bf 83} (1999) 3370--3373,
  [\href{http://arxiv.org/abs/hep-ph/9905221}{{\tt hep-ph/9905221}}].

\bibitem{Davoudiasl:1999jd}
H.~Davoudiasl, J.~Hewett, and T.~Rizzo, {\it {Phenomenology of the
  Randall-Sundrum Gauge Hierarchy Model}},  {\em Phys.Rev.Lett.} {\bf 84}
  (2000) 2080, [\href{http://arxiv.org/abs/hep-ph/9909255}{{\tt
  hep-ph/9909255}}].

\bibitem{Csaki:2005vy}
C.~Csaki, J.~Hubisz, and P.~Meade, {\it Tasi lectures on electroweak symmetry
  breaking from extra dimensions},
  \href{http://arxiv.org/abs/hep-ph/0510275}{{\tt hep-ph/0510275}}.

\bibitem{Gherghetta:2010cj}
T.~Gherghetta, {\it Tasi lectures on a holographic view of beyond the standard
  model physics},  \href{http://arxiv.org/abs/hep-ph/1008.2570}{{\tt
  hep-ph/1008.2570}}.

\bibitem{Dillon:2014zea}
B.~M. Dillon and S.~J. Huber, {\it {Non-Custodial Warped Extra Dimensions at
  the LHC?}},  {\em JHEP} {\bf 06} (2015) 066,
  [\href{http://arxiv.org/abs/1410.7345}{{\tt arXiv:1410.7345}}].

\bibitem{Ahmed:2015ona}
A.~Ahmed, B.~Grzadkowski, J.~F. Gunion, and Y.~Jiang, {\it {Higgs Dark Matter
  from a Warped Extra-Dimension -- the truncated-inert-doublet model}},  {\em
  JHEP} {\bf 10} (2015) 033, [\href{http://arxiv.org/abs/1504.0370}{{\tt
  arXiv:1504.0370}}].

\bibitem{Peskin:1991sw}
M.~E. Peskin and T.~Takeuchi, {\it Estimation of oblique electroweak
  corrections},  {\em Phys.Rev.} {\bf D46} (1992) 381--409.

\bibitem{Delgado:2007ne}
A.~Delgado and A.~Falkowski, {\it Electroweak observables in a general 5d
  background},  {\em JHEP} {\bf 0705} (2007) 097,
  [\href{http://arxiv.org/abs/hep-ph/0702234}{{\tt hep-ph/0702234}}].

\bibitem{Grossman:1999ra}
Y.~Grossman and M.~Neubert, {\it {Neutrino masses and mixings in
  nonfactorizable geometry}},  {\em Phys.Lett.} {\bf B474} (2000) 361--371,
  [\href{http://arxiv.org/abs/hep-ph/9912408}{{\tt hep-ph/9912408}}].

\bibitem{Gherghetta:2000kr}
T.~Gherghetta and A.~Pomarol, {\it A warped supersymmetric standard model},
  {\em Nucl.Phys.} {\bf B602} (2001) 3--22,
  [\href{http://arxiv.org/abs/hep-ph/0012378}{{\tt hep-ph/0012378}}].

\bibitem{Huber:2003tu}
S.~J. Huber, {\it Flavor violation and warped geometry},  {\em Nucl.Phys.} {\bf
  B666} (2003) 269--288, [\href{http://arxiv.org/abs/hep-ph/0303183}{{\tt
  hep-ph/0303183}}].

\bibitem{Baak:2012kk}
M.~Baak, M.~Goebel, J.~Haller, A.~Hoecker, D.~Kennedy, et~al., {\it The
  electroweak fit of the standard model after the discovery of a new boson at
  the lhc},  {\em Eur.Phys.J.} {\bf C72} (2012) 2205,
  [\href{http://arxiv.org/abs/hep-ph/1209.2716}{{\tt hep-ph/1209.2716}}].

\bibitem{Cabrer:2011fb}
J.~A. Cabrer, G.~von Gersdorff, and M.~Quiros, {\it Suppressing electroweak
  precision observables in 5d warped models},  {\em JHEP} {\bf 1105} (2011)
  083, [\href{http://arxiv.org/abs/hep-ph/1103.1388}{{\tt hep-ph/1103.1388}}].

\bibitem{Archer:2012qa}
P.~R. Archer, {\it {The Fermion Mass Hierarchy in Models with Warped Extra
  Dimensions and a Bulk Higgs}},  {\em JHEP} {\bf 09} (2012) 095,
  [\href{http://arxiv.org/abs/1204.4730}{{\tt arXiv:1204.4730}}].

\bibitem{Fichet:2013ola}
S.~Fichet and G.~von Gersdorff, {\it Anomalous gauge couplings from composite
  higgs and warped extra dimensions},  {\em JHEP} {\bf 1403} (2014) 102,
  [\href{http://arxiv.org/abs/hep-ph/1311.6815}{{\tt hep-ph/1311.6815}}].

\bibitem{Archer:2014jca}
P.~R. Archer, M.~Carena, A.~Carmona, and M.~Neubert, {\it {Higgs Production and
  Decay in Models of a Warped Extra Dimension with a Bulk Higgs}},
  \href{http://arxiv.org/abs/hep-ph/1408.5406}{{\tt hep-ph/1408.5406}}.

\bibitem{Ahmed:2015}
A.~Ahmed, B.~M. Dillon, and B.~Grzadkowski, {\it {Higgs-Radion Unification and
  Electroweak Precision Observables in Warped Extra Dimension}},  {\em in
  preparation}.

\bibitem{Peskin:2012we}
M.~E. Peskin, {\it Comparison of lhc and ilc capabilities for higgs boson
  coupling measurements},  \href{http://arxiv.org/abs/hep-ph/1207.2516}{{\tt
  hep-ph/1207.2516}}.

\bibitem{Asner:2013psa}
D.~Asner, T.~Barklow, C.~Calancha, K.~Fujii, N.~Graf, et~al., {\it Ilc higgs
  white paper},  \href{http://arxiv.org/abs/hep-ph/1310.0763}{{\tt
  hep-ph/1310.0763}}.

\bibitem{Gomez-Ceballos:2013zzn}
{\bf TLEP Design Study Working Group} Collaboration, M.~Bicer et~al., {\it
  First look at the physics case of tlep},  {\em JHEP} {\bf 1401} (2014) 164,
  [\href{http://arxiv.org/abs/hep-ex/1308.6176}{{\tt hep-ex/1308.6176}}].

\bibitem{Falkowski:2008fz}
A.~Falkowski and M.~Perez-Victoria, {\it Electroweak breaking on a soft wall},
  {\em JHEP} {\bf 0812} (2008) 107, [\href{http://arxiv.org/abs/0806.1737}{{\tt
  arXiv:0806.1737}}].

\bibitem{Carmona:2011ib}
A.~Carmona, E.~Ponton, and J.~Santiago, {\it Phenomenology of non-custodial
  warped models},  {\em JHEP} {\bf 1110} (2011) 137,
  [\href{http://arxiv.org/abs/hep-ph/1107.1500}{{\tt hep-ph/1107.1500}}].

\bibitem{Barbieri:2004qk}
R.~Barbieri, A.~Pomarol, R.~Rattazzi, and A.~Strumia, {\it {Electroweak
  symmetry breaking after LEP-1 and LEP-2}},  {\em Nucl.Phys.} {\bf B703}
  (2004) 127--146, [\href{http://arxiv.org/abs/hep-ph/0405040}{{\tt
  hep-ph/0405040}}].

\bibitem{CMS:2013xfa}
{\bf CMS Collaboration} Collaboration, {\it Projected performance of an
  upgraded cms detector at the lhc and hl-lhc: Contribution to the snowmass
  process},  \href{http://arxiv.org/abs/1307.7135}{{\tt arXiv:1307.7135}}.

\bibitem{ATLAS:2013hta}
{\it {Physics at a High-Luminosity LHC with ATLAS}},  {\em ATLAS Collaboration}
  (2013) [\href{http://arxiv.org/abs/hep-ex/1307.7292}{{\tt
  hep-ex/1307.7292}}].

\bibitem{Casagrande:2008hr}
S.~Casagrande, F.~Goertz, U.~Haisch, M.~Neubert, and T.~Pfoh, {\it {Flavor
  Physics in the Randall-Sundrum Model: I. Theoretical Setup and Electroweak
  Precision Tests}},  {\em JHEP} {\bf 0810} (2008) 094,
  [\href{http://arxiv.org/abs/hep-ph/0807.4937}{{\tt hep-ph/0807.4937}}].

\bibitem{Buras:2009ka}
A.~J. Buras, B.~Duling, and S.~Gori, {\it {The Impact of Kaluza-Klein Fermions
  on Standard Model Fermion Couplings in a RS Model with Custodial
  Protection}},  {\em JHEP} {\bf 0909} (2009) 076,
  [\href{http://arxiv.org/abs/hep-ph/0905.2318}{{\tt hep-ph/0905.2318}}].

\bibitem{Azatov:2009na}
A.~Azatov, M.~Toharia, and L.~Zhu, {\it {Higgs Mediated FCNC's in Warped Extra
  Dimensions}},  {\em Phys.Rev.} {\bf D80} (2009) 035016,
  [\href{http://arxiv.org/abs/hep-ph/0906.1990}{{\tt hep-ph/0906.1990}}].

\bibitem{Beneke:2014sta}
M.~Beneke, P.~Moch, and J.~Rohrwild, {\it {Muon anomalous magnetic moment and
  penguin loops in warped extra dimensions}},  {\em Int.J.Mod.Phys.} {\bf A29}
  (2014) 1444011, [\href{http://arxiv.org/abs/hep-ph/1404.7157}{{\tt
  hep-ph/1404.7157}}].

\end{thebibliography}\endgroup
\bibliographystyle{jhep}

\end{document}